\documentclass[dvips]{article}

\usepackage{icrc2011}

\title{Neutrino Solar Flare detection for a saving alert system of satellites and astronauts}

\shorttitle{D.Fargion , Solar Neutrino Flare Alert}

\authors{Daniele Fargion$^{1,2}$ }
\afiliations{$^1$Physics Department, Rome University 1, Sapienza,  Ple.A.Moro 2, Rome,Italy\\ $^2$INFN Rome 1 }
\email{daniele.fargion@Roma1.infn.it}

\abstract{Largest Solar Neutrino Flare may be soon detectable by Deep Core neutrino detector
immediately and comunicate to satellites or astronauts.  Its detection is the fastest
manifestation of a later (tens minutes,hours) dangerous cosmic shower. The precursor
trigger maybe saving satellites and even long flight astronauts lives. We shall
suggest how. Moreover their detection may probe the inner solar flare acceleration
place as well as the neutrino flavor mixing in a new different parameter windows. We
show the updated expected rate and signature of neutrinos and antineutrinos in
largest solar flare for present tens Megaton Deep Core telescope at tens Gev range.
Speculation for additional Icecube gigaton array signals are also considered. }
\keywords{ The keywords will be used to select your subject from all
ICRC contributions. }

\begin{document}
\maketitle
\section{Deep Core and Solar neutrino Flare}
During largest solar flare, of a few minutes duration,  the
particle flux escaping the corona eruption and hitting later on
the Earth, is 3-4 order of magnitude above the common atmospheric
CR background . If the flare particle interactions on the Sun corona is taking place as efficiently as  in terrestrial atmosphere, than their secondaries
by charged pions and muons decays, are leading to a neutrinos
fluency on Earth comparable to one day terrestrial atmospheric
neutrino activity (upper Bound). One therefore may expect a prompt increase of
neutrino signals of the order of one day integral events made by
atmospheric neutrinos \cite{ref0}. In present neutrino detectors the signal is
just on the edge, but as long as the authors know, it has been
never revealed \cite{ref1},\cite{ref4}. Sun density at the flare corona might be diluted
and pion production maybe consequently suppressed (by a factor
(0.1-0.05)) respect to terrestrial atmosphere, leading to a signal at few percent the expected 
 one under the above considerations. This may be the reason for the SK null detection. Indeed the low Gamma signals recently reported \cite{ref2} confirm this suppressed signal, but just at the detection edge (See Fig.\ref{01},\ref{02}). Unfortunately the neutrino signal at  hundred MeV energies is rare while the one at ten MeV or below is polluted by Solar Hep neutrinos.  The expected signal is dominated by $10-30$ MeV neutrinos, that  might be greatly improved  by   anti-neutrino component via Gadolinium presence in next SK detectors \cite{ref1}.  Our earliest (2006) upper limit
estimate for  October - November $2003$ solar flares \cite{ref0} and  the recent January $20th$ 2005 \cite{ref1} exceptional flare were  leading to signal
near unity for  Super-Kamiokande and to a few events above   unity for few Megaton detectors (See Fig.\ref{01},\ref{02}). Indeed large size  neutrino detectors ($5-10$ Megaton) at ten GeV (Deep Core)  are just recording data;
  ten GeV is a little high energy but hard solar flare have been observed up to tens GeV gamma. Moreover even GeV (PINGU) detectors, may soon be born.  Our recent estimate based on a lower (gamma) bound of neutrino signal, while below the upper ones, still confirms the order of magnitude and the near edge discovery (See Fig.\ref{01},\ref{02}). The recent peculiar solar flares as the October-November $2003$ and January $2005$ \cite{ref2} were source of high energetic charged particles: large fraction of these {\itshape{primary}} particles, became a source of both neutrons \cite{ref1} and {\itshape{secondary}} kaons, $K^{\pm}$, pions, $\pi^{\pm}$ by their particle-particle spallation on the Sun surface \cite{ref0}. Consequently,
$\mu^{\pm}$, final secondaries muonic and electronic neutrinos and anti-neutrinos,
${\nu}_{\mu}$, $\bar{\nu}_{\mu}$, ${\nu}_{e}$, $\bar{\nu}_{e}$,
$\gamma$ rays, are released by the chain reactions $\pi^{\pm}
\rightarrow \mu^{\pm}+\nu_{\mu}(\bar{\nu}_{\mu})$, $\pi^{0}
\rightarrow 2\gamma$, $\mu^{\pm} \rightarrow
e^{\pm}+\nu_{e}(\bar{\nu}_{e})+ \nu_{\mu}(\bar{\nu}_{\mu})$
occurring on the sun atmosphere. There are two different sites for
these decays (see \cite{ref0}): A brief and sharp
solar flare, originated within the $solar$ corona itself  and a diluted
and delayed $ terrestrial$ neutrino flux, produced by late flare
particles hitting the  Earth's  atmosphere. This latter delayed
signal is of poor physical interest, like an inverse missing signal during the E. Forbush  phase. The main and first {\itshape{solar}} flare neutrinos reach the Earth with a well defined directionality and within a narrow time range. The
corresponding average energies $<E_{{\nu}_{e}}>$,
$<E_{{\nu}_{\mu}}>$  ( since  low solar corona densities)
suffer negligible energy loss: $<E_{{\nu}_{e}}> $ $\simeq$ $50
MeV$, $<E_{{\nu}_{\mu}}> \simeq$ 100 $\div$ 200 MeV. The opposite
occur to downward flare. In the simplest approach, the main source
of pion production is $p+p\rightarrow {{\Delta}^{++}}n\rightarrow
p{{\pi}^{+}}n$; $p+p\rightarrow{{{\Delta}^{+}}p}^{\nearrow^{p+p+{{\pi}^{0}}}}_{\searrow_{p+n+{\pi}^{+}}}$
at the center of mass of the resonance ${\Delta}$ (whose mass
value is ${m}_{\Delta}=1232$ MeV). As  a first approximation and
as a useful simplification after the needed boost of the
 secondaries energies  one may assume that the total pion $\pi^+ $ energy
is equally distributed, in average, in all its final remnants:
($\bar{\nu}_{\mu}$, ${e}^{+}$, ${\nu}_{e}$,
${\nu}_{\mu}$):${E}_{{\nu}_{\mu}} \geq {E}_{{\bar{\nu}_{\mu}}}
\simeq {E}_{{\nu}_{e}} \simeq \frac{1}{4}{E}_{{\pi}^{+}}$. Similar
nuclear reactions (at lower probability) may also occur by
proton-alfa scattering leading to: $p+n\rightarrow
{{\Delta}^{+}}n\rightarrow n{{\pi}^{+}}n$; $p+n\rightarrow
{{{\Delta}^{o}}p}^{\nearrow^{p+p+{{\pi}^{-}}}}_{\searrow_{p+n+{\pi}^{o}}}$.
Here we neglect the ${\pi}^{-}$ additional role due to the flavor
mixing and the dominance of previous reactions ${\pi}^{+}$. To a
first approximation the flavor oscillation will lead to a
decrease in the muon component and it will make the electron
neutrino component a bit harder. Indeed the oscillation length (at
the energy considered) is small with respect to the Earth-Sun distance:
 $ L_{\nu_{\mu}-\nu_{\tau}}=2.48 \cdot10^{9} \,cm \left(
 \frac{E_{\nu}}{10^{9}\,eV} \right) \left( \frac{\Delta m_{ij}^2
 }{(10^{-2} \,eV)^2} \right)^{-1} \ll D_{\oplus\odot}=1.5\cdot
 10^{13}cm$.  While at the birth place the neutrino
fluxes by positive charged pions $\pi^+$ are
$\Phi_{\nu_e}$:$\Phi_{\nu_{\mu}}$:$\Phi_{\nu_{\tau}}$ $= 1:1:0$,
after the mixing assuming a democratic number redistribution we
expect $\Phi_{\nu_e}$:$\Phi_{\nu_{\mu}}$:$\Phi_{\nu_{\tau}}$ $=
(\frac{2}{3}):(\frac{2}{3}):(\frac{2}{3})$. Naturally in a more
detailed balance the role of the most subtle and hidden parameter
 (the very  recent, possibly  detected, neutrino mixing  $\Theta_{13}$ \cite{ref5})  may be deforming the present
averaged  flavor balance. On the other side for the anti-neutrino
fluxes we expect at the birth place:
$\Phi_{\bar\nu_e}$:$\Phi_{\bar\nu_{\mu}}$:$\Phi_{\bar\nu_{\tau}}$
$= 0:1:0$ while at their arrival (within a similar democratic
redistribution): $\Phi_{\bar\nu_e}$:$\Phi_{\bar\nu_{\mu}}$:$\Phi_{\bar\nu_{\tau}}$
$= (\frac{1}{3}):(\frac{1}{3}):(\frac{1}{3})$.
This neutrino  flux, derived by gamma one, hold $100$ s  duration and it is larger by
two order of magnitude over the atmospheric one.
\section{Solar Flare in Deep Core }
Therefore in present paper we conclude  that, even if SK just marginally missed the Solar neutrino flares (See Fig.\ref{01},\ref{02}),  Hyper Kamiokande, HK, or Megaton, Titand \cite{ref3} or at best Deep Core \cite{ref4}, and future PINGU  detectors should discover solar neutrino signals quite above threshold edge (See Fig.\ref{03}), \cite{ref4}. Finally full $km^3$ ICECUBE, being detecting at highest $50$ GeV may or may-not reveal tens GeV neutrinos (as observed gamma found in Milagro ones) just at edge (See Fig.\ref{04}).Solar Neutrino $\nu_e$, $\nu_{\mu}$ Flare After their
mixing in two different upper bounds considered in early papers \cite{ref0},\cite{ref1},\cite{ref4},derived by Solar flare energy equipartition  in two solar corona density target. Antineutrino exhibit  comparable  fluxes over   a noise-free solar antineutrino background \cite{ref1}. Here the signal is derived, by  pions connection and gamma detection of Solar flare on 20 January $2005$,\cite{ref2} in SK. While upper bounds appears  well within SK detection,  the observed lower gamma bound quite below, is just (marginally) out of the SK detection thresholds. This is consistent and it  explains the solar flare neutrino absence (SK private communication). The solar  flare duration and power (here and in next figures), from where we derived  the expected neutrino signals are assumed about $100$ s long  as powerful as the observed $20$ Jan. $2005$ event. The vertical arrows among the arcs describes our estimated solar flare neutrino flux windows. The vertical dotted lines are related to SK and Deep Core or PINGU cut-threshold via neutrino-nucleon  CC, whose cross-section are quite variables with energy. The solar
 neutrino noise rules lowest energies, but not above  tens MeV energy band. Icecube as a gigaton at hundred GeV maybe not ideal.
 In figures above,  the  dashed line , along $\nu_{\mu}$ , shows the muon neutrino thresholds,(near GeV), just out SK detection but within Deep Core and PINGU ranges. The different dotted lines split on from  electrons, taus and the most inclined  ones for muons because cross section energy depandance. In conclusion a Deep Core alarm system for satellites and astronauts maybe the fastest and most useful ones. Astronauts on the longest flight and largest spacestation may find a partial screening by  hiding  each other in largest volume  water reserve containing (one or few  tons of water) in full immersion, just  a few hour after the solar neutrino flare.


\begin{thebibliography}{99}
\bibitem{ref0}{D.Fargion,JHEP$06$,(2004),045; D.Fargion,F.Moscato, Chin.J.Astron.Astrophys.3(2003)S75-S76.}
\bibitem{ref1}{D.Fargion,Phys.Scripta,T127,(2006),22-24}
\bibitem{ref2}{Grechnev V.V. et al.,arXiv:0806.4424, Solar Physics,Vol.1,October,(2008),252}
 \bibitem{ref3}{Matthew D. Kistler et al. 0810.1959v1}
\bibitem{ref4} {D.Fargion,Nuclear Physics B (Proc. Suppl.) 188 (2009) 142–145}
\bibitem{ref5}{K2K Collaboration, K. Abe et al.(15 June 2011) $ www.t2k.org/docs/pub/003/t2k-nue1st.pdf $}
 
\vspace{-20pc}
\begin{figure}
\input epsf
\includegraphics[width=9cm,height=7cm]{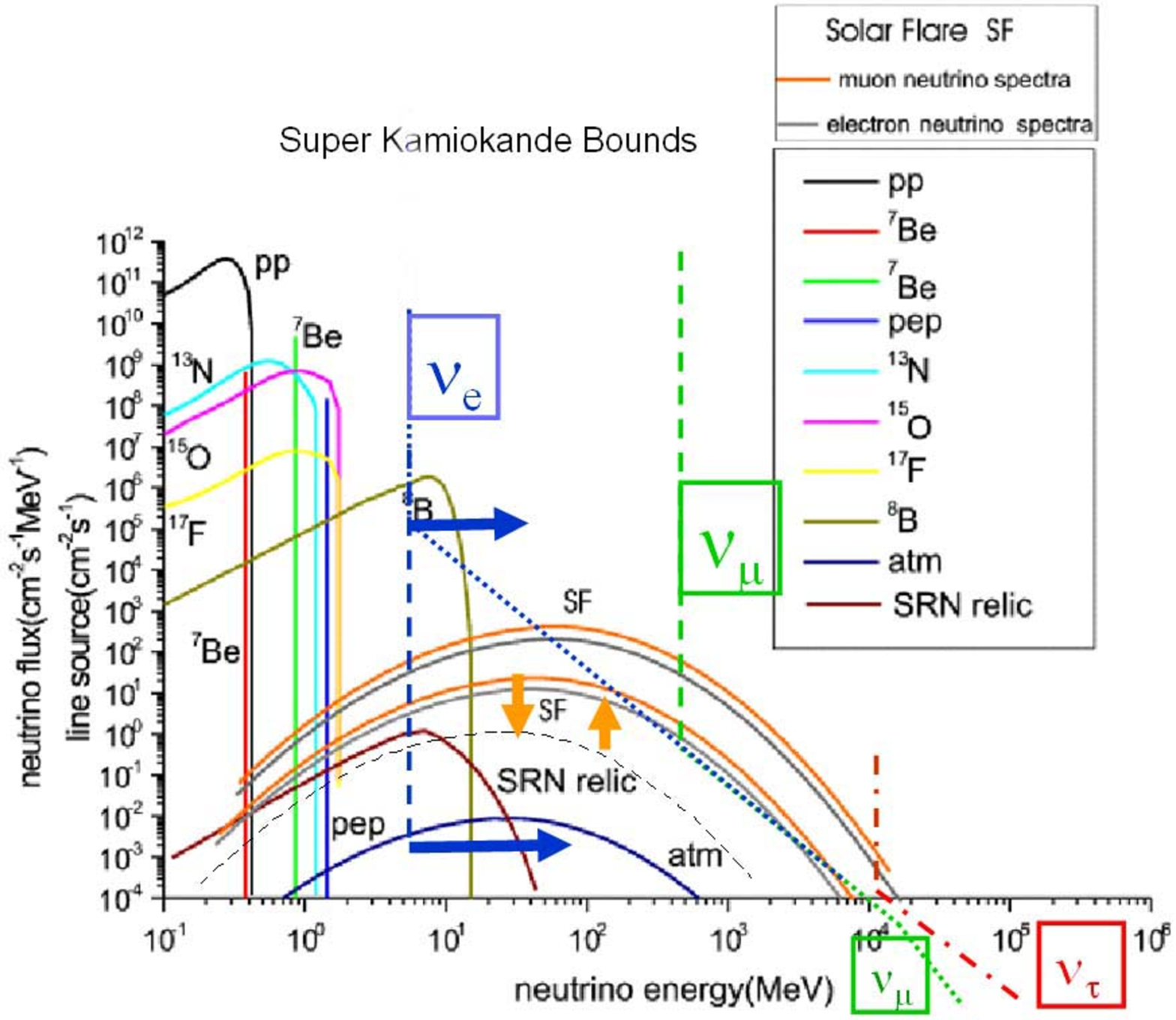}
\caption{Solar Neutrino $\nu_e$, $\nu_{\mu}$ flare versus noises. Note the expected windows by up-down arrows.}\label{01}

\input epsf
\includegraphics[width=9cm,height=7cm]{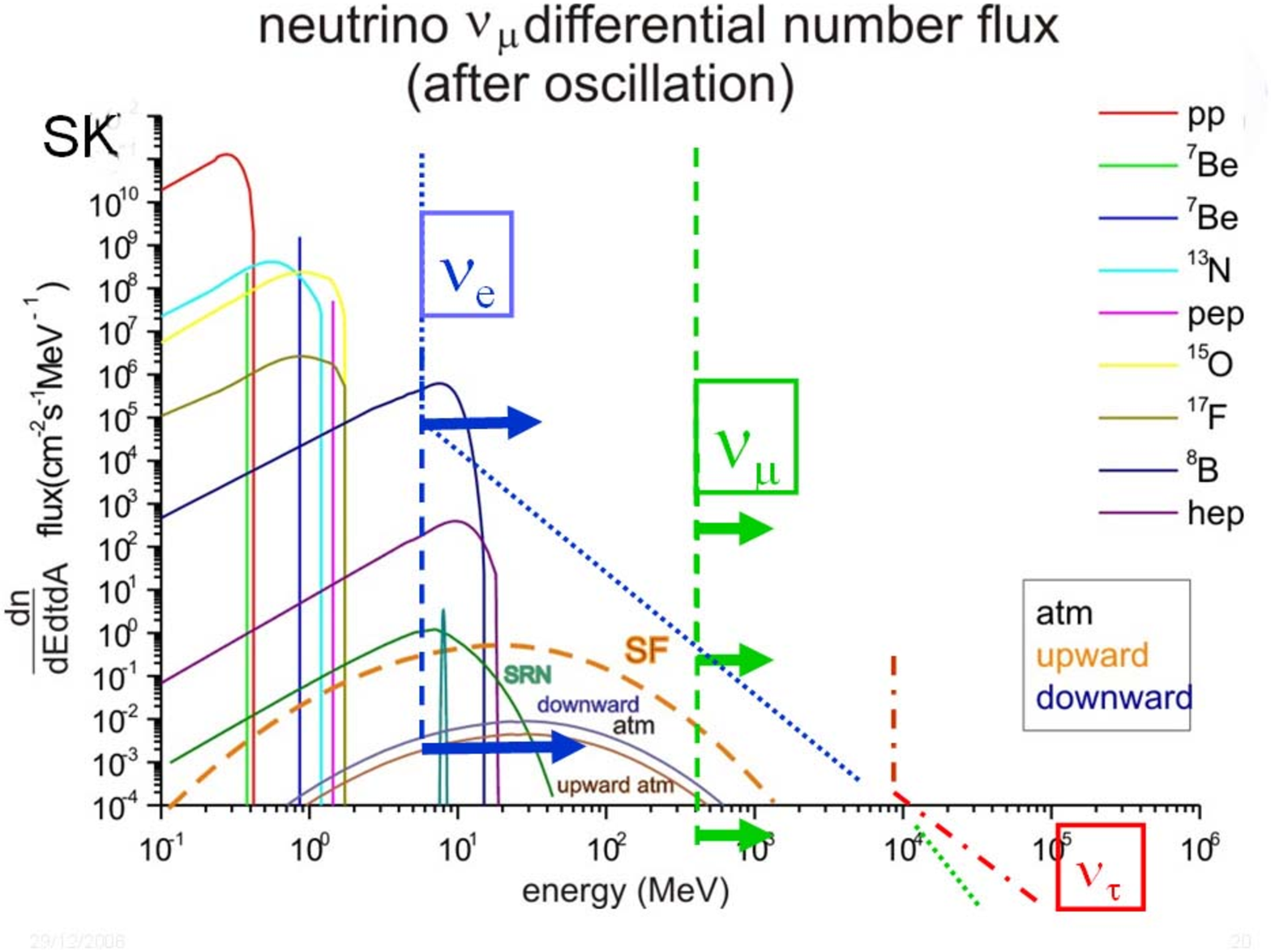}
\caption{ As above  in SK, by  $\gamma$ flux calibration impling  a Solar $\nu_e$, $\nu_{\mu}$ neutrino flare as small as $5\%$ of the corresponding cosmic solar flare, because a much diluted solar atmosphere. }\label{02}
\end{figure}



\begin{figure}
\includegraphics[width=9cm,height=8cm]{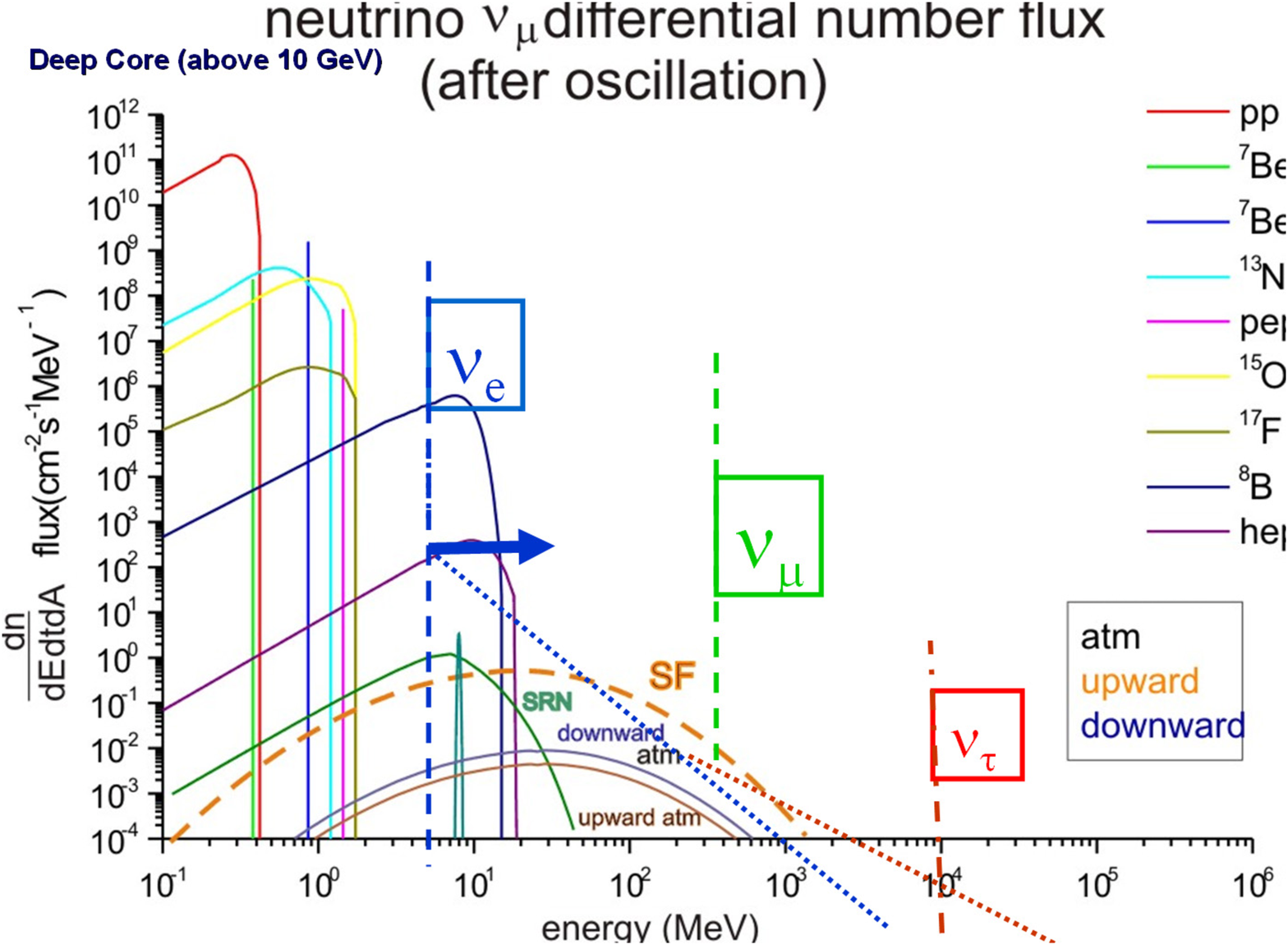}
\caption{As above, for Deep Core detecotor above  10 GeV. As shown the  $\nu_{\mu}$ signal is nearly revealed at the GeV edge; however future PINGU enhanced detector might reach the threshold of GeV neutrino discover from solar flare }\label{03}

\end{figure}\label{04}


\end{thebibliography}
\end{document}